\documentstyle[aaspptwo]{article}
\begin{document}

\title{EVOLUTION OF CLUSTER AND FIELD ELLIPTICALS AT $0.2 < z < 0.6$
IN THE {\em CNOC} CLUSTER SURVEY}

\author{David Schade\footnotemark[1], R. G. Carlberg\footnotemark[1], H.K.C. Yee\footnotemark[1] and Omar L\'{o}pez-Cruz\footnotemark[2]\footnotemark[3]} 
\affil{Department of Astronomy, University of Toronto, Toronto, Canada M5S 3H8}
\author{Erica Ellingson\footnotemark[1]}
\affil{Center for Astrophysics and Space Astronomy, University of Colorado, CO 80309, USA}

\footnotetext[1]{Visiting Astronomer, Canada-France-Hawaii Telescope,
which is operated by the National Research Council of Canada, the Centre
Nationale de la Recherche Scientifique of France, and the University
of Hawaii.}

\footnotetext[2]{{\it also} Instituto Nacional de Astrof\'{\i}sica Optica y Electr\'onica
(INAOE), Tonantzintla, M\'exico}

\footnotetext[3]{Visiting Astronomer, Kitt Peak 
National Observatory, National Optical Astronomy Observatories, which
is operated by the Association of Universities for Research in
Astronomy, Inc., under cooperative agreement with the National Science
Foundation.}

\begin{abstract}

Two-dimensional surface photometry has been done for 166
{\em early-type} galaxies  
(bulge/total luminosity $B/T>0.6$) in 3 fields of the
Canadian Network for Observational Cosmology (CNOC) cluster survey.
These galaxies are either spectroscopically confirmed members of clusters at 
$z=0.23$ (45 galaxies), $0.43$ (22), and $0.55$ (16) or field
galaxies in the same redshift range.  
An additional 51 early-type galaxies in the rich cluster Abell 2256 at
$z=0.06$  were analysed with the same technique. The resulting
structural and surface brightness measurements show that,
in the plane of absolute  magnitude $M_{AB}(B)$ versus $\log R_e$ (half-light radius),
the locus of cluster ellipticals
shifts monotonically with redshift so that at redshifts of (0.23, 0.43, 0.55), galaxies
of a given size are more luminous by
($-0.25\pm 0.10,-0.55\pm 0.12,-0.74\pm 0.21$) magnitudes with respect to
the same relation measured at $z=0.06$ (adopting $q_\circ=0.5$). 
There is  no evidence that
early-type galaxies in the field evolve differently from those
in clusters.
If dynamical processes do not substantially modify
the size-luminosity relation for early-type galaxies over the observed redshift
range, then these galaxies have undergone significant
luminosity evolution over the past half of the age of the
universe. The amount of brightening is consistent with passive
evolution models of old, single-burst stellar populations.

\end{abstract}

\keywords{galaxies:evolution---galaxies:fundamental parameters}

\section{INTRODUCTION}

 Luminosity evolution of early-type galaxies has long been predicted to occur
as an inevitable consequence of an aging stellar population
(Tinsley 1972) but its detection has been proven difficult.
Dressler and Gunn (1990) found signs of color evolution 
(expected to accompany luminosity evolution) among even
the reddest cluster galaxies by $z\sim 0.7$. More
recent studies of distant clusters (Aragon-Salamanca et al.; 1993
and Rakos and Schombert; 1995)
report changes in color with redshift which are
broadly consistent with early-forming
and passively-evolving elliptical galaxy models (e.g., Bruzual 1993).
  Yee and Green (1987) found an apparent 
brightening of the characteristic magnitude of 
the luminosity function of cluster galaxies associated with quasars by $0.9 \pm 0.5$
mag at $z=0.6$.

  Observations with {\em Hubble Space Telescope} have opened up
new opportunities for morphological studies of cluster
galaxies. Pahre, Djorgovski, \& de Carvalho (1996) find evolution of
$0.36 \pm 0.14$ magnitudes in the restframe $K$ band from Early-Release
Observations of Dressler et al. (1994) of the cluster CL0939+4713 (Abell 851) at $z=0.41$.
The same $HST$ imaging was used by Barrientos, Schade, \& L\'{o}pez-Cruz (1996) to
derive a value of $0.64\pm 0.3$ mag of evolution from $z=0.41$ to the present.
These values are consistent with passively evolving models
of elliptical galaxies (Tinsley 1972, Bruzual 1993).
Thus it appears that luminosity evolution of early-type galaxies has
been directly detected in one cluster. It is important
to establish whether this observation is representative of the early-type 
population as a whole.

 Although {\em HST} resolution ($\sim 1 $ kpc at $z > 0.5$) is necessary 
to resolve bars, dust lanes, and spiral structure in galaxies
at high redshift, it has been shown (Schade et al. 1996) that
ground-based imaging is capable of providing quantitative measurements
of the gross morphology of distant galaxies. In particular, it was found that disk scale
lengths ($h$) and bulge effective radii ($R_e$) can be measured reliably
under certain conditions. Disk scale length can be usefully measured in
mid- to late-type galaxies (fractional bulge luminosity $B/T < 0.5$),
whereas elliptical/bulge effective radii can be reliably measured for
early-type objects ($B/T > 0.7$). The fractional bulge luminosity
itself was shown to be measurable with a dispersion of $\sim 20\%$
(Schade et al 1996) in faint ($I_{AB} \sim 22$) galaxies for high-redshift
objects in the Canada-France Redshift Survey (CFRS). The typical size
of CFRS galaxies is $h \sim R_e \sim 0.35 ^{\prime \prime}$.  

The Canadian Network for Observational Cosmology (CNOC)  cluster survey (Carlberg et al. 1994,
Yee, Ellingson, \&
Carlberg 1996)
provides a unique dataset for the study of the evolution of
cluster and field galaxies. In addition to good redshift coverage
(16 clusters with $0.2 < z < 0.6$), the CNOC survey contains
redshifts for 2600 galaxies all the way from the cluster core
($R_c < 0.5$ Mpc) to the low-density outer regions ($R_c \sim 3$ Mpc)
and into the field. Abraham et al. (1996) exploited this
wide range in environments in a study of the galaxy populations
in A2390.
Particularly relevant to the present work is
the fact that directly comparable samples of field galaxies
at each redshift are available to complement the cluster samples.
The typical galaxy size in the present work is disk scale
length $\sim$ bulge effective radius $\sim 0.7 ^{\prime \prime}$
so that the ratio of typical size to seeing (FWHM) is 0.7, compared
to a ratio of 0.5 in Schade et al. (1996) for CFRS ground-based
imaging. 

This {\it Letter}  concentrates on the morphological analysis of 124 
cluster and 66 field {\em early-type} galaxies at $0.06 < z < 0.55$ representing
the first phase of a comprehensive study of the evolution of
cluster galaxies and cluster/field differential evolution. 
Observations, and procedure  are  described in \S 2. 
The relation between size and luminosity
or surface brightness are presented in \S 3 and the
results are discussed in \S 4.
 It is assumed throughout this paper that $H_\circ=50$ km sec$^{-1}$ Mpc$^{-1}$ 
and $q_{\circ}=0.5$.

\section{OBSERVATIONS AND PROCEDURE}

 Imaging was obtained in June and October 1993 
using the Canada-France-Hawaii Telescope  Multi-Object
Spectrograph (MOS).  Gunn $r$-band  imaging
was used to fit the two-dimensional luminosity
distributions in this analysis. 
Integration times were 900 seconds for A2390 (Yee et. al 1996) 
and MS1621+26 (Ellingson et al. 1996)
and 1200 seconds for MS0016+16. 
These three clusters were chosen
to yield a good range in redshift (0.228, 0.427 and 0.547) and
to contain reasonably large numbers of
spectroscopically confirmed cluster members (174, 98, and 47,
respectively). The MOS image quality was fairly good for these clusters
with seeing of 
$0.93\pm0.06$ and $0.97\pm0.04$ arcseconds (FWHM)  for the central and
inner east fields in A2390, $1.04\pm 0.04$ 
and  $1.2 \pm 0.04$ for
the two fields in MS1621 (central and south), and, $1.00\pm 0.04$ in MS0015. These dispersions are 
from gaussian fits to azimuthally averaged stellar profiles
and represent the variation of the PSF {\em core} over the regions of the frame
where fitted galaxies are located.  An empirical point-spread function (PSF)
for each frame was constructed using DAOPHOT routines (Stetson 1987).

  The analysis procedure was identical to that described by Schade et al.
(1995). Galaxy parameters (size, surface brightness,
and fractional bulge luminosity, $B/T$) were estimated by
constructing ``symmetrized''  images of the galaxies 
(see Schade et al. 1995) in the Gunn $r$-band.
The use of images that are symmetric by construction minimizes the effects of nearby
companions and other irregular structure. These images
were fit with  two-dimensional galaxy models integrated over each pixel
and convolved with the empirical point-spread functions.
The two components
used are an exponential disk and a deVaucouleurs ($R^{1/4}$) 
law. The majority of local galaxies have luminosity distributions that are
well-described by some combination of these components (Kormendy 1977, Kent 1985, 
Kodaira, Watanabe,
and Okamura 1986)
and  {\em Hubble Space Telescope} work confirms that this is
also true for high-redshift galaxies (Schade et al. 1995, Barrientos
et al. 1996).

  A set of 561 galaxies in these three CNOC fields with velocities,
regardless of cluster membership or color, were 
subjected to the two-dimensional fitting procedure. Fitting
failed to converge for 22 of the galaxies (4\%), usually
because of close neighbors and image defects. Failures were
discarded. 

 In addition to the CNOC clusters, fits were done on 100 bright galaxies  
on an 1800 second B image of the cluster A2256 from the L\'{o}pez-Cruz \& Yee
survey (L\'{o}pez-Cruz 1996) obtained with the Kitt Peak 0.9 meter telescope. 
These galaxies were chosen to be within $\pm 0.1$ magnitudes
in $B-R$ of the tight red cluster galaxy sequence in the color-magnitude
diagram thus ensuring
a high probability of both cluster membership and early-type morphology.
  Fits were done using an identical procedure to that used for
the CNOC galaxies. 

 Those objects with a bulge fraction $B/T > 0.6$ as measured from the best-fit
two-dimensional models were defined as early-type galaxies. After applying
this selection criterion, the median values
of $B/T$ for the clusters A2256, 
A2390, 
MS1621+26, and MS0016+16 are 0.81, 0.83, 0.83, 0.94
respectively. The CNOC field samples had similar median
$B/T$ values. The numbers of early-type galaxies for these clusters
repsectively are 51, 45, 22, and 16.
In all cases, the pure-bulge model fit values of $M_{AB}(B)$ and $R_e$ were adopted.
The adoption of pure bulge model parameters (as opposed to bulge-plus-disk parameters)
may result in overestimates of the bulge size in the presence of  a
typical disk component. The size of such an effect will be very small
given the large median values of $B/T$ in the samples.
The observed Gunn $g$ and $r$ magnitudes
and colors were converted to restframe $M_{AB}(B)$ luminosities and
$(U-V)_{AB,\circ}$ colors [$(U-V)_{AB}=(U-V)+0.7$] based on interpolation
among the spectral energy distributions of Coleman, Wu, and Weedman (1980)
as described by Lilly et al. (1995). The galaxies
in A2256 were also K-corrected according to Coleman, Wu, and Weedman (1980).

\section{RESULTS}

Figure 1 shows  the relation between $R_{e}$ (half-light radius) and
luminosity for cluster and field {\em early-type} galaxies  at
$0.06 < z < 0.55$.  We measure the change
in the galaxy loci with redshift assuming they can be represented simply by
shifts along the luminosity axis. The similarity of the slopes in the individual
panels tends to support this approach but no physical interpretation is
necessarily implied. A slope
of $\Delta M/ \Delta log R_E=-3.33$ (the mean of fits to the cluster and field
galaxy loci individually, including the Coma fit from Barrientos et al 1996)
was adopted for the cluster and field galaxies.
The magnitude shifts $\Delta M$ were estimated using a constrained linear fit
using this fixed slope and the errors are given by $s/\sqrt{(n-1)}$ where $s$ is
the estimated dispersion and $n$ is the number of data points. 

 Superimposed on Figure 1 are the best-fit lines to the galaxy loci for each
cluster and these {\em cluster } loci are plotted on the field galaxy
panels (these are {\em not } the best-fit field galaxy lines). 
 The best-fit shifts in luminosity for the clusters and
 field along with their uncertainties are given in Table 1.
 The data in Figure 1 and Table 1 show that cluster galaxies of a given size 
grow progressively more luminous with increasing redshift. 
The corresponding amount of
 brightening in the field galaxy sample is consistent with that in the clusters
 at similar redshift. 
 The galaxy luminosity  enhancement is well-described by: $\Delta M_B=-1.35 z$
and this is equivalent to an increase in surface brightness (at a given 
size) by this amount. 

 The effect of cosmology on this result is indicated by the arrows in the lower
left of each cluster panel. These show the change in size and luminosity that
result from changing $q_\circ=0.5$ to $q_\circ=0.1$. The net effect
on the  computed magnitude shifts relative to the cluster A2256 is an increased
evolution by $-0.02,-0.05$, and $-0.09$ for the clusters at $z=0.23,0.43$ and $0.55$
respectively.

\section{DISCUSSION}

 Two conclusions follow from the present observations.  First, 
the relationship between  $M_B$ and $\log R_e$ for early-type
cluster galaxies shifts progressively with redshift such that  by $z=0.55$
a galaxy of a given size is more luminous by $-0.74\pm 0.21$ mag than
its counterpart at $z=0.06$. In other words, the surface brightness
has increased by this amount. 
 Second, evolution of the $M_B-\log R_e$ relation for early-type population 
in the field is also observed 
(see Table 1) and the amount of
brightening is consistent with that observed in clusters at similar
redshift. Thus, there is no indication from this study that early-type
galaxies in clusters evolve differently than those 
in the field environment. It is important to note, however, that our sample of
cluster galaxies is dominated by those far (up to several Mpc) from the
high-density cluster core.

If the size-luminosity relation in clusters and in the field
is universal (so that the comparison done here between high-redshift and
local galaxies is valid) and if 
dynamical evolution does not significantly change 
the structure of early-type galaxies over this
redshift range (i.e., the sizes remain constant), then we are 
seeing luminosity evolution of individual galaxies. A similar
amount of evolution ($\Delta M^* =-0.2,-0.5$, 
and $-0.5$ at $z=0.2,0.4$, and $0.6$
respectively) was detected in the luminosity function of
galaxies in clusters associated with quasars by Yee and Green (1987)
and is consistent with preliminary results of an analysis of the CNOC cluster luminosity
function (Yee et al. in preparation, Crete
proceedings).
Barrientos et al (1996) found evolution of $\Delta M_B
=-0.6 \pm 0.3$ mag in the cluster CL0939+4713 and Bender,R. Ziegler, B.,
\& Bruzual, G. 1996) derive a similar amount of evolution
($\sim 0.5$ mag) from velocity dispersions and Mg absorption
line strengths for 16 elliptical galaxies in the cluster 
MS1512+36 at $z\sim0.37$ (part of the CNOC survey).

 The $M_B-\log R_e$ relation is a projection of the fundamental plane
of elliptical galaxies (Djorgovski \& Davis 1987) whose properties
have been found to vary between cluster and field samples (de Carvalho \& Djorgovski 1992)
in a number of respects, with field ellipticals representing a
less homogeneous population than those in clusters. Although
we see no sign of that effect in the present sample, we cannot exclude
it and it is important to consider this issue in detail in future work. Dynamical evolution
of ellipticals could complicate the interpretation of morphological results
such as those presented here although simulations (Capelato, de Carvalho, \&
Carlberg 1995) indicate that dynamical evolution may simply change an objects'
position  on the fundamental plane rather than modifying the position of
the plane itself. In this case no evolution of the fundamental pane would occur with redshift
except that due to the evolution of the stellar populations themselves.

 If interpreted as luminosity evolution of individual galaxies,
 the amount of brightening measured in the present study is consistent
with that expected for a single-burst stellar population formed
at high redshift (see Figure 2).
Models published by Bruzual (1993) and Buzzoni (1995)
predict a  brightening of $\Delta M_B=-0.6$ to $-1.2$
magnitudes between $z=0$ and $z=0.6$ for a single-burst population with an
age of 15 Gyr. These results are remarkably similar to those obtained
by Tinsley (1972) and represent a gratifying confirmation of an
effect predicted 25 years ago.
The exact amount of brightening 
depends strongly on the initial mass function (IMF) and less strongly on
the age of the population and cosmology.
Taken at face value, the data
presented here agree better with models based on IMFs with 
proportionally more high-mass stars and flatter power-law mass
functions than the standard Salpeter (power-law index 2.35) 
initial mass function.

 This work represents the first phase of a comprehensive analysis of
the cluster and field populations in the CNOC survey. A number of
questions that are beyond the scope of the present work clearly need to be answered.
For example, Figure 1 shows a range of  scatter about the mean relation. The
cluster MS1621 has a much smaller scatter than the other clusters and it 
is important to know whether this is due to observational error or 
is a reflection of intrinsic
differences between clusters. A complete analysis of the variation
of galaxy properties with distance from the cluster core and with redshift is the fundamental 
goal of the CNOC morphology project. It is also important to reconcile these results
with analyses of the luminosity functions in this survey (Yee et al. in preparation,
Lin et al. in preparation) and in other surveys (Lilly et al. 1995). A comparative
study of the disk galaxy population in clusters and the field,
similar to that done here
for ellipticals, is clearly important.

\begin{acknowledgments}

 We thank Simon Lilly, Felipe Barrientos, and Pedro Colin for helpful discussions. 
 This work was supported financially by NSERC of Canada.

\end{acknowledgments}

\newpage
\makeatletter
\def\jnl@aj{AJ}
\ifx\revtex@jnl\jnl@aj\let\tablebreak=\nl\fi
\makeatother


\begin{planotable}{ccccccc}
\tablewidth{30pc}
\tablecaption{Evolution of the $M_B-\log R_e$ relation for CNOC galaxies}
\tablehead{
\colhead{Cluster}           & \colhead{z}      &
\colhead{$\Delta M_B$} & \colhead{N}       & \colhead{Field }  &
\colhead{$\Delta M_B$} &  \colhead{N} }
\startdata
Abell 2390 & 0.228  & $-0.25\pm 0.10$ & 40 & $0.12 < z < 0.32$ & $-0.2 \pm 0.2$ & 14 \nl
MS1621+26 & 0.427  & $-0.55\pm 0.12$ & 19 & $0.32 < z < 0.52$ & $-0.8 \pm 0.2$ &  33 \nl
MS0016+16 & 0.547  & $-0.74\pm 0.21$ & 16 & $0.45 < z < 0.65$ & $-0.8 \pm 0.2$ &  15 \nl
\tablecomments{N gives the number of galaxies with $M_B(AB)<-20$ that were used to
derive the values of $\Delta M_B$ given in this table. These results assume $q_\circ=0.5$
and the effect of cosmology appears in the text.}

\end{planotable}

\newpage
\centerline{Figure captions}

\begin{figure}
\caption{ The relation between $M_{AB}(B)$ and $\log R_e$ (half-light
radius in kpc) for early-type galaxies (measured $B/T > 0.6$). Clusters
are shown in the left panels and corresponding field samples
on the right. The best-fit fixed-slope relation from the cluster A2256
is superimposed (solid line) on each of the panels. The best-fit
relation for each cluster is also plotted (dotted lines) and
this {\em cluster} line is superimposed on the corresponding field galaxy 
panel. All fits were restricted to $M_{AB}(B) < -20$.
The differences between the best-fit cluster and field
relations are not statistically significant.}

\end{figure}

\begin{figure}
\caption { The luminosity shift $\Delta M$  from the
$M_B-\log R_e$ relation is plotted against redshift. 
Also shown are the theoretical tracks for the passive evolution
of a single-burst stellar population formed 15 Gyr before the present
time (with $\Omega_\circ=0.5$)
from Buzzoni (1995) 
for three values ($s=1.35$,2.35, and 3.35) of the power-law index $s$ of a Salpeter
initial mass function (IMF). The giant-rich IMF ($s=1.35$) produces
the largest amount of evolution. 
The models of Bruzual (1993) predict a
flatter slope and larger amount of evolution at $0 < z  < 0.3$
but are also consistent with these observations.}
\end{figure}

\end{document}